\documentclass[%
    reprint,
    superscriptaddress,
    nofootinbib, nobibnotes,
    amsmath,amssymb, aps, prd,
    floatfix, ]{revtex4-2}

\usepackage[english]{babel}

\usepackage{acro}
\usepackage{amsmath}
\usepackage[caption=false]{subfig}
\usepackage{booktabs}
\usepackage{graphicx}% Include figure files
\usepackage{dcolumn}% Align table columns on decimal point
\usepackage{bm}% bold math
\usepackage[dvipsnames]{xcolor}
\definecolor{myblue}{RGB}{62,73,173}
\definecolor{myred}{RGB}{216, 28, 56}
\usepackage[colorlinks=true,breaklinks=true,hyperfootnotes=true,allcolors=myblue]{hyperref}
\usepackage[nameinlink]{cleveref}

\Crefname{section}{Sec.}{Sec.}
\Crefname{figure}{Fig.}{Figs.}
\Crefname{table}{Tab.}{Tabs.}

\usepackage{physics}
\usepackage{siunitx}
\AtBeginDocument{\RenewCommandCopy\qty\SI}

\renewcommand{\qc}{\, \text{,}}
\newcommand{\qs}{\, \text{.}}
\DeclareAcroEnding{possessive}{'s}{'s}
\NewAcroCommand\acg{m}{\acropossessive\UseAcroTemplate{first}{#1}}
\NewAcroCommand\acsg{m}{\acropossessive\UseAcroTemplate{short}{#1}}
\NewAcroCommand\aclg{m}{\acropossessive\UseAcroTemplate{long}{#1}}
\NewAcroCommand\acfg{m}{%
    \acrofull
    \acropossessive
    \UseAcroTemplate{first}{#1}%
}
\NewAcroCommand\iacsg{m}{%
    \acroindefinite
    \acropossessive
    \UseAcroTemplate{short}{#1}%
}
\DeclareAcronym{AK}{
    short = AK,
    long  = analytic kludge
}
\DeclareAcronym{AAK}{
    short = AAK,
    long  = argumented analytic kludge
}
\DeclareAcronym{AUC}{
    short = AUC,
    long  = area under the curve
}
\DeclareAcronym{BH}{
    short = BH,
    long  = black hole
}
\DeclareAcronym{BHB}{
    short = BHB,
    long  = black hole binary,
    long-plural-form = black hole binaries
}
\DeclareAcronym{BBH}{
    short = BBH,
    long  = binary black hole
}
\DeclareAcronym{CNN}{
    short = CNN,
    long  = convolutional neural network
}
\DeclareAcronym{DECIGO}{
    short = DECIGO,
    long  = DECi-hertz Interferometer Gravitational wave Observatory
}
\DeclareAcronym{DECODE}{
    short = DECODE,
    long  = DilatEd COnvolutional neural network for Detecting Extreme-mass-ratio inspirals
}
\DeclareAcronym{DNN}{
    short = DNN,
    long  = deep neural network
}
\DeclareAcronym{ESA}{
    short = ESA,
    long  = European Space Agency
}
\DeclareAcronym{EMRI}{
    short = EMRI,
    long  = extreme-mass-ratio inspiral
}
\DeclareAcronym{FPR}{
    short = FPR,
    long  = false positive rate
}
\DeclareAcronym{GB}{
    short = GB,
    long  = galactic binary,
    long-plural-form = galactic binaries
}
\DeclareAcronym{GR}{
    short = GR,
    long  = general relativity
}
\DeclareAcronym{GW}{
    short = GW,
    long  = gravitational wave
}
\DeclareAcronym{LDC}{
    short = LDC,
    long  = LISA Data Challenge
}
\DeclareAcronym{LIGO}{
    short = LIGO,
    long  = \href{http://www.ligo.caltech.edu/}{Laser Interferemeter Gravitational Wave Observatory}
}
\DeclareAcronym{LISA}{
    short = LISA,
    long  = \href{https://www.lisamission.org/}{Laser Interferometer Space Antenna}
}

\DeclareAcronym{LSO}{
    short = LSO,
    long  = last stable orbit
}

\DeclareAcronym{MBH}{
    short = MBH,
    long  = massive black hole
}
\DeclareAcronym{MBHB}{
    short = MBHB,
    long  = massive black hole binary,
    long-plural-form = massive black hole binaries
}

\DeclareAcronym{MCMC}{
    short = MCMC,
    long  = Markov-chain Monte Carlo
}
\DeclareAcronym{MLDC}{
    short = MLDC,
    long  = \href{http://astrogravs.nasa.gov/docs/mldc/}{Mock LISA Data Challenge}
}
\DeclareAcronym{NK}{
    short = NK,
    long  = numerical kludge
}
\DeclareAcronym{OMS}{
    short = OMS,
    long  = optical metrology system
}
\DeclareAcronym{PSD}{
    short = PSD,
    long  = power spectral density
}
\DeclareAcronym{ReLU}{
    short = ReLU,
    long  = rectified linear
    unit
}

\DeclareAcronym{ROC}{
    short = ROC,
    long  = receiver operating characteristic
}
\DeclareAcronym{SGWB}{
    short = SGWB,
    long  = stochastic gravitational wave background
}
\DeclareAcronym{SMBH}{
    short = SMBH,
    long  = super-massive black hole
}
\DeclareAcronym{SNR}{
    short = SNR,
    long  = signal-to-noise ratio
}
\DeclareAcronym{SOBH}{
    short = SOBH,
    long  = stellar origin black hole binary
}
\DeclareAcronym{SSB}{
    short = SSB,
    long  = solar system barycenter
}
\DeclareAcronym{TCN}{
    short = TCN,
    long  = temporal convolutional network
}
\DeclareAcronym{TDI}{
    short = TDI,
    long  = time delay interferometry
}
\DeclareAcronym{TPR}{
    short = TPR,
    long  = true positive rate
}
\DeclareAcronym{t-SNE}{
    short = t-SNE,
    long  = t-distributed stochastic neighbor embedding
}
\DeclareAcronym{VGB}{
    short = VGB,
    long  = verification galactic binary,
    long-plural-form = verification galactic binaries
}

\begin{document}

\preprint{APS/123-QED}

\title{Dilated convolutional neural network for detecting extreme-mass-ratio inspirals}

\author{Tianyu Zhao}
\affiliation{Department of Astronomy, Beijing Normal University, Beijing 100875, China}
\affiliation{Peng Cheng Laboratory, Shenzhen, 518055, China}
\affiliation{Institute for Frontiers in Astronomy and Astrophysics, Beijing Normal University, Beijing 102206, China}

\author{Yue Zhou}
\affiliation{Peng Cheng Laboratory, Shenzhen, 518055, China}

\author{Ruijun Shi}%
\affiliation{Department of Astronomy, Beijing Normal University, Beijing 100875, China}
\affiliation{Institute for Frontiers in Astronomy and Astrophysics, Beijing Normal University, Beijing 102206, China}

\author{Zhoujian Cao}%
\thanks{Corresponding author: \href{mailto:zjcao@bnu.edu.cn}{zjcao@bnu.edu.cn}}
\affiliation{Department of Astronomy, Beijing Normal University, Beijing 100875, China}
\affiliation{Institute for Frontiers in Astronomy and Astrophysics, Beijing Normal University, Beijing 102206, China}
\affiliation{School of Fundamental Physics and Mathematical Sciences, Hangzhou Institute for Advanced Study, UCAS, Hangzhou 310024, China}

\author{Zhixiang Ren}
\thanks{Corresponding author: \href{mailto:renzhx@pcl.ac.cn}{renzhx@pcl.ac.cn}}
\affiliation{Peng Cheng Laboratory, Shenzhen, 518055, China}

\date{\today}% It is always \today, today,
%  but any date may be explicitly specified

\begin{abstract}
    % 1. current issue
    The detection of Extreme Mass Ratio Inspirals (EMRIs) is intricate due to their
    complex waveforms, extended duration, and low signal-to-noise ratio (SNR),
    making them more challenging to be identified compared to compact binary
    coalescences.
    % 2. existing method
    While matched filtering-based techniques are known for their computational
    demands, existing deep learning-based methods primarily handle time-domain data
    and are often constrained by data duration and SNR. In addition, most existing
    work ignores time-delay interferometry (TDI) and applies the long-wavelength
    approximation in detector response calculations, thus limiting their ability to
    handle laser frequency noise.
    % 3. our method
    In this study, we introduce DECODE (\acl{DECODE}), an end-to-end model focusing
    on EMRI signal detection by sequence modeling in the frequency domain.
    % 4. our innovation
    Centered around a dilated causal convolutional neural network, trained on
    synthetic data considering TDI-1.5 detector response, DECODE can efficiently
    process a year's worth of multichannel TDI data with an SNR of around 50.
    % 5. our results
    % 5.1 (SNR, duration, accuracy, speed)
    We evaluate our model on 1-year data with accumulated SNR ranging from 50 to
    120 and achieve a true positive rate of 96.3\% at a false positive rate of 1\%,
    keeping an inference time of less than 0.01 seconds.
    % 5.2 (science and future)
    With the visualization of three showcased EMRI signals for interpretability and
    generalization, DECODE exhibits strong potential for future space-based
    gravitational wave data analyses.
\end{abstract}

% \keywords{Suggested keywords}
% Use showkeys class option if keyword
% display desired

\maketitle

\section{\label{sec:intro} Introduction}
% GW discovery and space-based GW detection
The groundbreaking detection of \acp{GW} in 2015, exemplified by the GW150914
event, has profoundly impacted the field of astrophysics \cite{abbott_2016}.
Enabled by the \ac{LIGO} \cite{the_ligo_scientific_collaboration_2015} and
Virgo \cite{acernese_2015}, this remarkable achievement unequivocally confirmed
the existence of \acp{GW}, providing empirical validation of \ac{GR}
\cite{abbott_2016b}. Beyond enriching our knowledge of the cosmos, this seminal
discovery has ushered in a new era of astronomical observation
\cite{bailes_2021}. With the spotlight now turning to space-based \ac{GW}
observatories \cite{amaro-seoane_2023,sesana_2016a}, the absence of terrestrial
disturbances allows for a more dedicated exploration of the low-frequency
\acp{GW} \cite{matichard_seismic_2015}. This exciting pursuit carries the
potential to reveal hitherto unobserved phenomena, offering profound insights
into the nature of our universe \cite{bailes_2021}.

% the gravitational physics and astrophysics perspective of EMRI signals and it is the core scientific objective of space-based detection
Space-based \ac{GW} detection, a largely unexplored domain, marks the next
epoch in astrophysics \cite{amaro-seoane_2023}. Pioneering this exciting
venture are projects such as the \ac{LISA} \cite{amaro-seoane_2017} by the
\ac{ESA}, with NASA's participation, and Asian projects including Japan's
\ac{DECIGO} and B-DECIGO \cite{kawamura_2019,kawamura_2021}, as well as China's
Taiji \cite{hu_2017,ren_2023} and TianQin \cite{luo_2016} missions. Targeting
the millihertz frequency band, these endeavors offer a novel perspective for
the exploration of diverse astrophysical and cosmological phenomena through the
detection of low-frequency \acp{GW}
\cite{amaro-seoane_2023,auclair_2023,barausse_2020}. The scientific goals are
broad, with the intent to shed light on the enigmas of \acp{MBHB}, \acp{EMRI},
continuous waves from \acp{GB}, and the stochastic \ac{GW} backgrounds produced
by the early universe's myriad of unresolved sources \cite{scird_2018}.

\begin{figure*}[!ht]
    \centering
    \subfloat[\label{fig:td-data}]{%
        \includegraphics[width=0.62\textwidth]{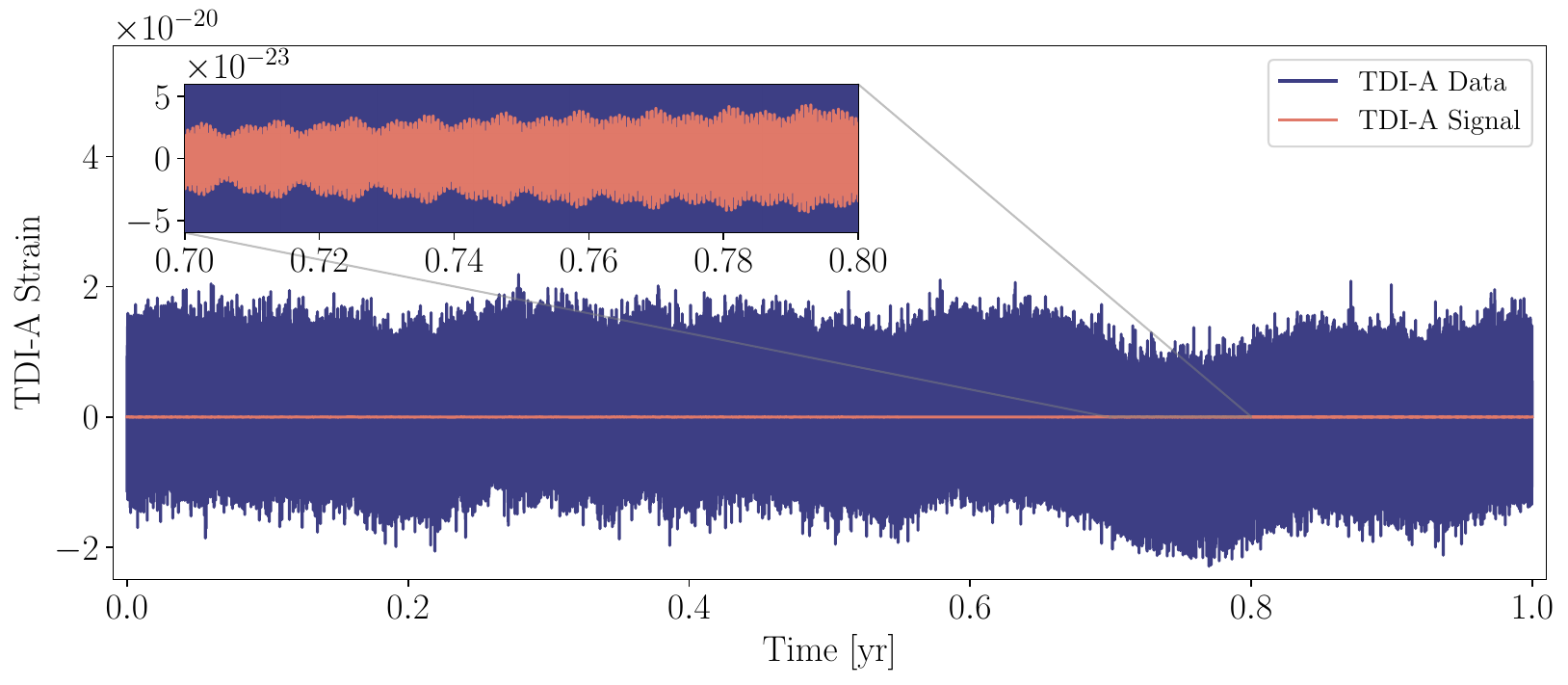}
    }
    \hfill
    \subfloat[\label{fig:fd-data}]{%
        \includegraphics[width=0.36\textwidth]{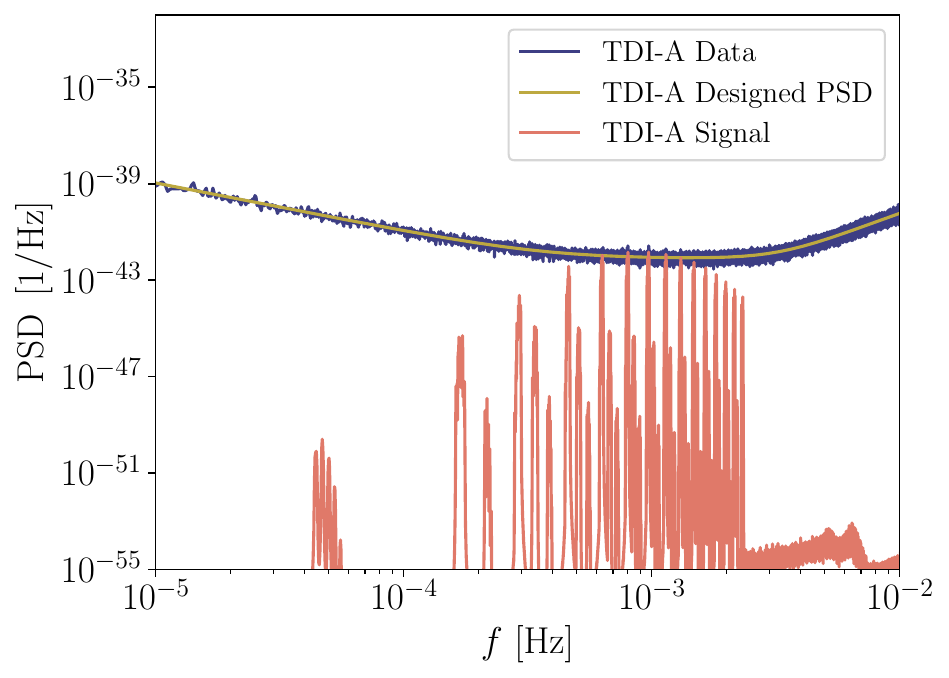}
    }
    \caption{\textbf{Visualization of a training data sample.} This depicts an \ac{EMRI} signal from the \acs{TDI}-A channel spanning 1-year with an \acs{SNR} of 70.  (a), Time-domain representation of the TDI-A strain, showcasing both the combined data (signal + noise) and the signal. The signal's amplitude is about 3 orders of magnitudes lower than the noise, which makes the detection challenging. (b), Welch \acs{PSD} of the combined data and the signal, the signal contains lots of modes (peaks), with some reaching the noise level, highlighting the suitability of the frequency domain detection method. The designed detector noise \acs{PSD} is also presented for reference.}
    \label{fig:data}
\end{figure*}

In the spectrum of potential discoveries, \acp{EMRI} hold a unique position.
These events, initiated when a compact stellar remnant spirals into a \ac{MBH},
provide opportunities to investigate the \ac{MBH} characteristics and the
nature of the surrounding environments \cite{gair_2017}. \acp{EMRI} emit
low-frequency \acp{GW} throughout their extended inspiral phase, serving as a
rich source of information for understanding system physical parameters and the
\acg{MBH} spacetime geometry \cite{babak_2017}. The successful detection and
parameter estimation of \ac{EMRI} signals could provide novel insights into the
astrophysics of \acp{MBH} and the foundational principles of gravity
\cite{chua_2018,han_2019}. Traditional methods for \ac{EMRI} detection, which
include both time-domain and time-frequency-domain techniques, have been widely
studied in prior research \cite{gair_2004a,gair_2005,
    wen_2005,gair_2007,gair_2008}. These strategies mainly employ matched filtering
\cite{gair_2004a,gair_2007} and the Short Time Fourier Transform
\cite{gair_2005,wen_2005,gair_2008}. However, the inherent complexities of
\ac{EMRI} signals present significant obstacles. Characterized by their complex
waveform templates, high-dimensional parameter space, and multiple modes within
a single waveform, \ac{EMRI} signals require over $\sim 10^{35}$ templates for
matched filtering search \cite{babak_2017}, resulting in a computationally
intensive and time-consuming procedure. An example of single \ac{EMRI} in both
the time and frequency domains can be seen in \Cref{fig:data}, showcasing the
aforementioned challenges of signal detection. Additionally, \ac{EMRI} signals
are typically faint and buried within detector and confusion noise,
necessitating extended observation durations to achieve an adequate \ac{SNR}
for detection \cite{babak_2017}. Time-frequency techniques, offering
representations in both time and frequency domains, are frequently less
sensitive than matched filtering, which limits their ability to identify weak
signals \cite{gair_2008}. Given these challenges, exploring alternative
methods, such as deep learning, becomes crucial for potentially improving the
efficiency of \ac{EMRI} signal detection.

Deep learning, an advanced branch of machine learning, employs neural networks
with multiple layers for different types of data. By facilitating the
extraction of intricate patterns and representations from large datasets, it
has played a crucial role in advancing various fields, from image recognition
\cite{Lecun_1998} to natural language processing \cite{devlin_2018}. Among the
numerous architectures, the \ac{CNN} stands out for its proficiency in handling
structured data, such as images and time series, by progressively learning
features in a hierarchical manner. Starting with simple features like edges in
the initial layers, they gradually combine these to recognize more complex
patterns and structures in the deeper layers. This layered approach allows
\acp{CNN} to automatically recognize and represent intricate details in the
data, making them highly effective for tasks like object detection
\cite{Redmon_2016} and time-series classification \cite{woo_2023}. In the area
of \ac{GW} data analysis, the potential of deep learning, especially \acp{CNN},
is becoming increasingly evident. A large amount of studies
\cite{george_2018,gabbard_2018,wang_2020,krastev_2020, lopez_2021,
    skliris_2022,qiu_2023,ravichandran_2023} have demonstrated their effectiveness
in ground-based \ac{GW} detection. Beyond signal detection, deep learning
methods have been applied to a variety of tasks, including glitch
classification \cite{colgan_2020, cavaglia_2019, razzano_2018}, denoising
\cite{wei_2020,ren_2022,zhao_2023}, and parameter estimation
\cite{gabbard_2022,dax_2021}. However, the application of these methods to
space-based \ac{GW} detection is still in its early stages. While there have
been some exploratory efforts, such as the adoption of MFCNN \cite{wang_2020}
to detect \ac{MBHB} contaminated by confusion noise \cite{ruan_2023} and the
application of dictionary learning to low-\ac{SNR} space-based \ac{BBH}
detection \cite{Badger_2023}. Notably, Zhang et al. \cite{zhang_2022} pioneered
the detection of \acp{EMRI} using \ac{CNN}, though without incorporating the
\ac{TDI} technique. Therefore, further research is needed to harness the full
capabilities of deep learning in space-based GW analysis.

\begin{figure*}[!ht]
    \centering
    \subfloat[\label{fig:flowchart}]{%
        \includegraphics[width=0.99\textwidth]{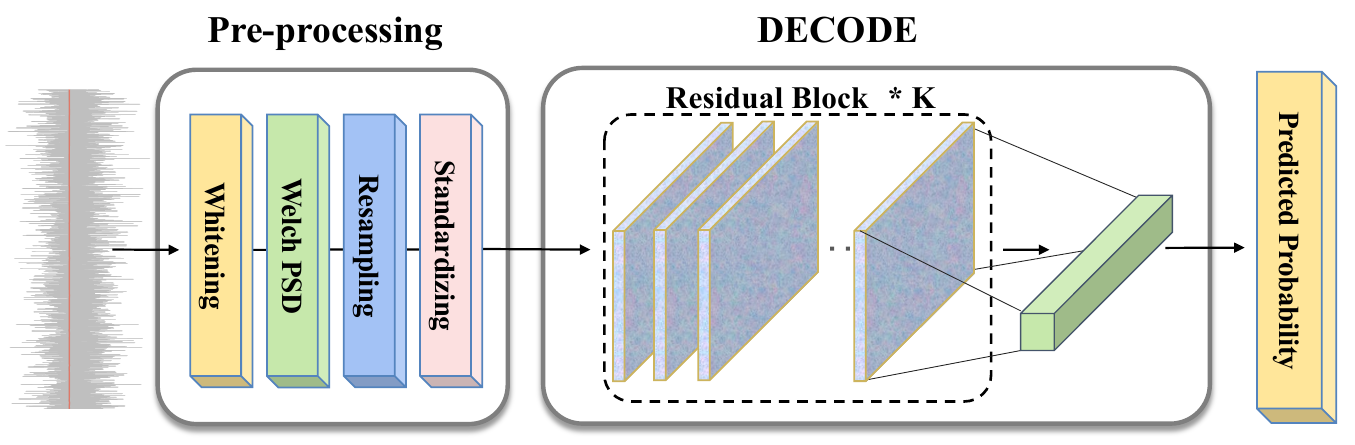}
    }\\
    \subfloat[\label{fig:wavenet}]{%
        \includegraphics[width=0.7\textwidth]{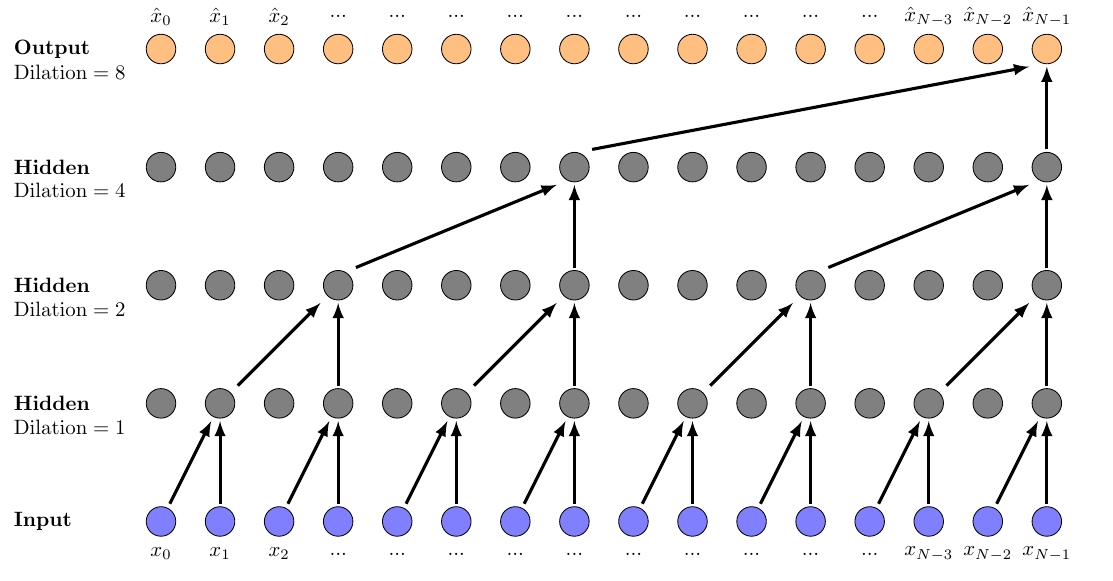}
    }
    \hfill
    \subfloat[\label{fig:res-block}]{%
        \includegraphics[width=0.28\textwidth]{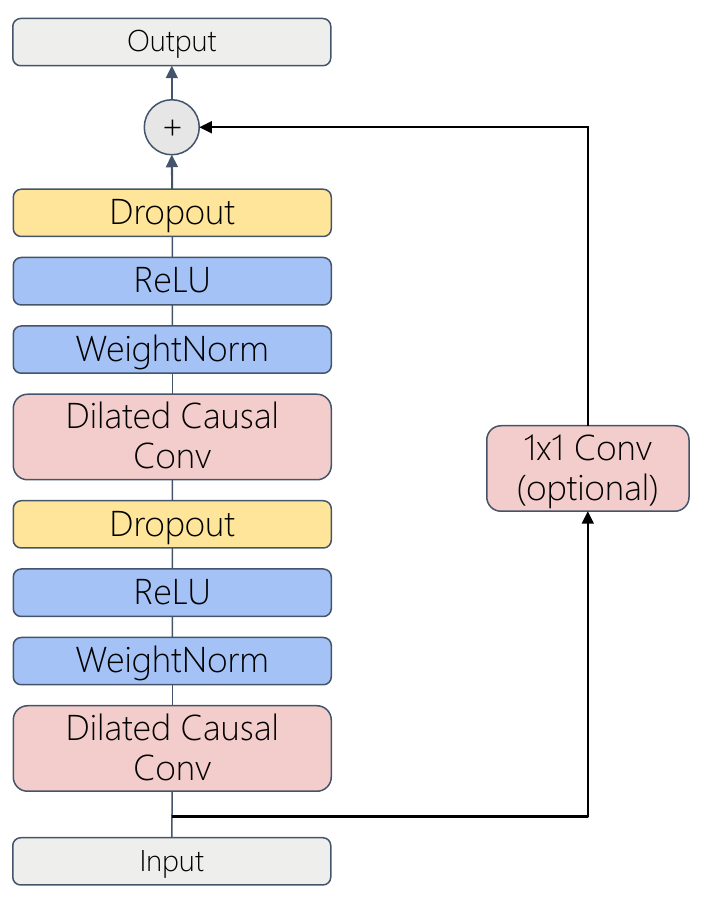}
    }
    \caption{\textbf{Comprehensive \ac{EMRI} detection framework.} (a), Depicts the entire \ac{EMRI} detection process, from initial data preprocessing to the end-to-end \acs{DECODE} model. (b), Highlights the mechanism of dilated causal convolution with dilation factors of $(1, 2, 4, 8)$ and a kernel size of 2, emphasizing the exponential growth of the receptive field. (c), Detailed architecture of the residual block in \acs{DECODE}, comprising two dilated causal convolutional layers, weight normalization, ReLU, and dropout layers. A $1\times1$ convolution is introduced to address any dimension discrepancies between the residual input and output.}
    \label{fig:network}
\end{figure*}

In this paper, we introduce the \acs{DECODE} (\acl{DECODE}), an end-to-end
model designed for detecting \ac{EMRI} signals in the frequency domain with an
\ac{SNR} of around 50. As showed in \Cref{fig:network}, the model incorporates
dilated causal convolutional layers, which expand its receptive field, allowing
it to efficiently process data covering an entire year in one pass. We trained
our model using synthetic data that considers the \ac{TDI}-1.5 detector
response, accounting for unequal arm lengths. The results are promising: the
\ac{DECODE} detects \ac{EMRI} signals with a 1-year accumulated \ac{SNR}
between 50 and 120, achieving a \ac{TPR} of 96.3\% with a \ac{FPR} of 1\%.
Notably, our model can evaluate one batch of data samples within $10^{-2}$
seconds. Visualizations of the model's intermediate outputs highlight its
interpretable feature extraction process and its ability to generalize beyond
\ac{GR}. These findings emphasize the potential of \ac{DECODE} in future
space-based \ac{GW} data analysis.

The remainder of this paper is organized as follows: \Cref{sec:method} provides
a detailed overview of the data generation procedure and outlines the
architecture of our proposed model, the \ac{DECODE}. In \Cref{sec:result}, we
present the results of our \ac{EMRI} detection experiments, demonstrating the
effectiveness of our approach. Finally, \Cref{sec:discussion} concludes the
paper with a summary of our findings and a discussion on potential future work
based on our findings.

\section{\label{sec:method} Method}
\subsection{EMRI Waveform Modeling}
% difficulties and importance of EMRI waveform modeling
% Existing EMRI template, their assumptions and difference
Detecting \acp{EMRI} has the potential to reveal key astrophysical insights,
but modeling their waveform is challenging due to the delicate balance of
strong-field \ac{GR} and gravitational radiation dynamics. Accurately
describing \acp{EMRI} demands a solution to the self-force problem, which
considers the gravitational impact of the smaller compact object on its own
motion within the powerful gravitational field of the central \ac{MBH}
\cite{amaro-seoane_2023}. Because the self-force problem is highly non-linear
and defies analytical solutions, researchers have developed approximate
waveform models, commonly referred to as kludge models
\cite{barack_2004,babak_2007}.

Two commonly used kludge models in \ac{EMRI} modeling are the \ac{AK}
\cite{barack_2004} model and the \ac{NK} \cite{babak_2007} model. The \ac{AK}
model relies on post-Newtonian expansions and perturbative calculations to
evolve the orbital parameters and generate waveforms quickly. It provides
computational efficiency but suffers from dephasing compared to more accurate
models, leading to potential inaccuracies in parameter estimation. On the other
hand, the \ac{NK} model incorporates the orbital trajectory computed in curved
space using Kerr geodesics and includes radiation reaction effects. Although
more accurate, the \ac{NK} model is computationally more expensive, making
\ac{EMRI} signal detection using this template highly formidable.

To address the limitations of both models, an \ac{AAK}
\cite{chua_2015,chua_2017,chua_2021} model has been proposed. The \ac{AAK}
model combines the computational efficiency of the \ac{AK} model with improved
phasing achieved through a mapping to Kerr geodesic frequencies and
self-consistent post-Newtonian evolution. By incorporating self-force
information and refining the phasing, the \ac{AAK} model achieves higher
waveform fidelity compared to the \ac{AK} model while remaining computationally
efficient. While its computational efficiency may not be adequate for matched
filtering-based signal searches, it is suitable for producing training datasets
for \acp{DNN}.

Despite the advancements in kludge waveform modeling, challenges remain.
Incorporating second-order self-force effects into the models and refining them
for orbits approaching plunge are ongoing areas of research
\cite{amaro-seoane_2023}. Nonetheless, these waveform models are crucial for
accurately representing the dynamics of \acp{EMRI} and enabling the detection,
parameter estimation, and data analysis of these elusive astrophysical sources.

\subsection{Data Curation}
The process of curating training and testing datasets for the identification of
\ac{EMRI} signals using a \ac{DNN} is a multi-step procedure consisting of
signal generation, detector response simulation, and pre-processing.

\paragraph{Waveform Generation}
The first step involves the generation of signal templates. The \ac{AAK} model
used for generating these templates is based on Ref. \cite{katz_2021}. The
waveform, denoted as $h(t) = h_+(t) - ih_\times(t)$, is typically characterized
by 14 physical parameters. The parameter space used for sampling the training
and testing dataset parameters in this study is detailed in
\Cref{tab:emri_par}. Here, $M$ and $a$ represent the mass and the spin
parameter of the \ac{MBH} respectively. The semi-latus rectum is denoted by
$p$, while $e$ stands for orbital eccentricity, and $\iota$ signifies the
orbit’s inclination angle from the equatorial plane. $Y = \cos \iota \equiv
    L_z/\sqrt{L_z^2+Q}$, where $Q$ is the Carter constant, and $L_z$ is the $z$
component of the specific angular momentum. For the orbital parameters $p$,
$e$, and $Y$, the initial values are designated as $p_0$, $e_0$, and $Y_0$
respectively. The polar and azimuthal sky location angles are represented by
$\theta_S$, and $\phi_S$. The orientation of the spin angular momentum vector
of the \ac{MBH} is described by the azimuthal and polar angles $\theta_K$ and
$\phi_K$. These parameters are uniformly sampled for our dataset. It is
important to note that the mass of the compact object, denoted by $\mu$ is
fixed at $10M_\odot$ and the initial phases for the azimuthal
($\Phi_{\varphi,0}$), polar ($\Phi_{\theta,0}$), and radial ($\Phi_{r,0}$)
modes are all manually set to 0 respectively.

\begin{table}[htbp]
    \begin{center}
        \caption{Summary of parameter setups in \ac{EMRI} signal simulation.}\label{tab:emri_par}%
        \begin{tabular}{@{}lcc@{}}
            \toprule
            \hline
            \textbf{Parameter}     & \textbf{Lower bound} & \textbf{Upper bound} \\
            \hline\hline
            $\log_{10}(M/M_\odot)$ & $5$                  & $8$                  \\
            $a$                    & $10^{-3}$            & $0.99$               \\
            $e_0$                  & $10^{-3}$            & $0.8$                \\
            $p_0/M$                & $15$                 & $25$                 \\
            $Y_0$                  & $-1$                 & $1$                  \\
            SNR                    & $50$                 & $120$                \\
            $\theta_S$             & $0$                  & $\pi$                \\
            $\phi_S$               & $0$                  & $2\pi$               \\
            $\theta_K$             & $0$                  & $\pi$                \\
            $\phi_K$               & $0$                  & $2\pi$               \\
            \hline
            \bottomrule
        \end{tabular}
    \end{center}
\end{table}

\begin{table}[htbp]
    \begin{center}
        \caption{Summary of configurations of training and testing dataset.} \label{tab:detector_par}%
        \begin{tabular}{@{}lr@{}}
            \toprule
            \hline
            \textbf{Parameter}                  & \textbf{Configuration}                                    \\
            \hline\hline
            Size of training dataset            & 5000                                                      \\
            Size of testing dataset             & 1000                                                      \\
            Cadence                             & \qty{15}{\second}                                         \\
            Duration                            & 1 year                                                    \\
            Re-sampled data length $N$          & 1024/2048/4096                                            \\
            \midrule
            Arm length $L$                      & $\qty{2.5e9}{\metre}$                                     \\
            Detector orbit                      & 1st order Keplerian orbit                                 \\
            TDI                                 & `     TDI-1.5                                             \\
            Acceleration noise $A_\mathrm{acc}$ & $\qty[per-mode=symbol]{3}{\femto\metre\per\sqrt{\hertz}}$ \\
            OMS noise $A_\mathrm{oms}$          & $\qty[per-mode=symbol]{15}{\pico\metre\per\hertz}$        \\
            \hline
            \bottomrule
        \end{tabular}
    \end{center}
\end{table}

\paragraph{TDI Response}
The next stage involves simulating the detector's response to these signals.
The specific detector configurations utilized in this study are detailed in
\Cref{tab:detector_par}. For the breathing arm length, we employed the
\ac{TDI}-1.5 technique, which yielded the \ac{GW} strain of \ac{TDI} A and E
channels, denoted as $h_A(t)$ and $h_E(t)$, respectively. A detailed derivation
of this technique can be found in Ref. \cite{katz_2022a}. Their CUDA-based
implementation, enable us to calculate the response cost in seconds. The signal
is then rescaled according to the desired \ac{SNR} using the formula:
\begin{align}
    \label{eq:snr}
    \mathrm{SNR}^2 = \left(h_A\mid h_A\right) + \left(h_E\mid h_E\right) \qs
\end{align}
Here, the inner product $(a\mid b)$ is defined as:
\begin{equation}
    (a \mid b) = 2\int_{f_{\mathrm{min}}}^{f_{\mathrm{max}}} \frac{\tilde{a}^*(f)\tilde{b}(f)+\tilde{a}(f)\tilde{b}^*(f)}{S_n(f)}\, \dd{f} \qs
\end{equation}
In this equation, $f_{\mathrm{min}} = \frac{1}{ \text{Duration}} \simeq \qty{3.17e-8}{\hertz}$ and $f_{\mathrm{max}} = \frac{1}{2\cdot\text{Cadence}} = \frac{1}{30} \unit{\hertz}$. $\tilde{a}(f)$ and $\tilde{b}(f)$ represent the frequency domain signals, and the superscript $*$ denotes the complex conjugate. $S_n(f)$ is the one side noise \ac{PSD}, which will be specified later.

\paragraph{Noise Generation}
The third step introduces noise to the signal. This noise is modeled as a
colored Gaussian noise with a \ac{PSD} defined by

\begin{equation}
    \mathrm{S_n}(f)=16 \sin ^2(\omega L)\left(P_{\mathrm{oms}}(f)+(3+\cos (2 \omega L)) P_{\mathrm{acc}}(f)\right)
    \qc
\end{equation}
with
\begin{equation}
    \begin{aligned}
        P_{\mathrm{oms}}(f) & = A_\mathrm{oms}^2 \left[1+\left(\frac{\qty{2}{\milli\hertz}}{f}\right)^4\right]\left(\frac{2 \pi f}{c}\right)^2 \qc \\
        P_{\mathrm{acc}}(f) & = A_\mathrm{acc}^2
        \left[ 1+\left(\frac{\qty{0.4}{\milli\hertz}}{f}\right)^2\right]                                                                           \\
                            & \quad \cdot\left[1+\left(\frac{f}{\qty{8}{\milli\hertz}} \right)^4 \right] \left(\frac{1}{2 \pi fc}\right)^2 \qs
    \end{aligned}
\end{equation}

Where $A_\mathrm{acc}$ and $A_\mathrm{oms}$ are the noise budget of test mass
acceleration noise and readout noise coming from the \ac{OMS}, $\omega = 2\pi
    f/c$, $L$ is the arm length of \ac{LISA} detector, and $c$ is the speed of
light. Then the signal is injected into the noise, resulting in the synthetic
data \Cref{fig:data} is a showcase of the training data in the time and
frequency domain.

\paragraph{Whitening and PSD Estimation}
In the final stage of data curation, the data undergoes several pre-processing
steps to prepare it for input into the \ac{DNN}. The first of these steps is
whitening, which serves to remove the frequency-dependent variations in the
noise. This process allows the \ac{DNN} to concentrate on the underlying signal
patterns, simplifying the learning task and enhancing the network's ability to
detect subtle patterns in the data, thereby improving the overall performance
of the \ac{EMRI} signal identification. Following whitening, the \ac{PSD} of
the data is estimated using Welch's method. The data then undergoes
sub-sampling, where it is re-sampled onto a log-uniform frequency grid. This
step is aimed at reducing the computational load of subsequent analyses by
decreasing the number of data points. 3 different grid density is selected as
listed in \Cref{tab:detector_par}. The final pre-processing step is
standardization, which ensures that all input features are on a uniform scale,
a fundamental requirement for most deep learning algorithms. This step is
crucial in enhancing the learning efficiency of the neural network and
improving the overall performance of the model.

\begin{figure*}[htbp]
    \subfloat[\label{fig:roc1}]{%
        \includegraphics[width=0.32\textwidth]{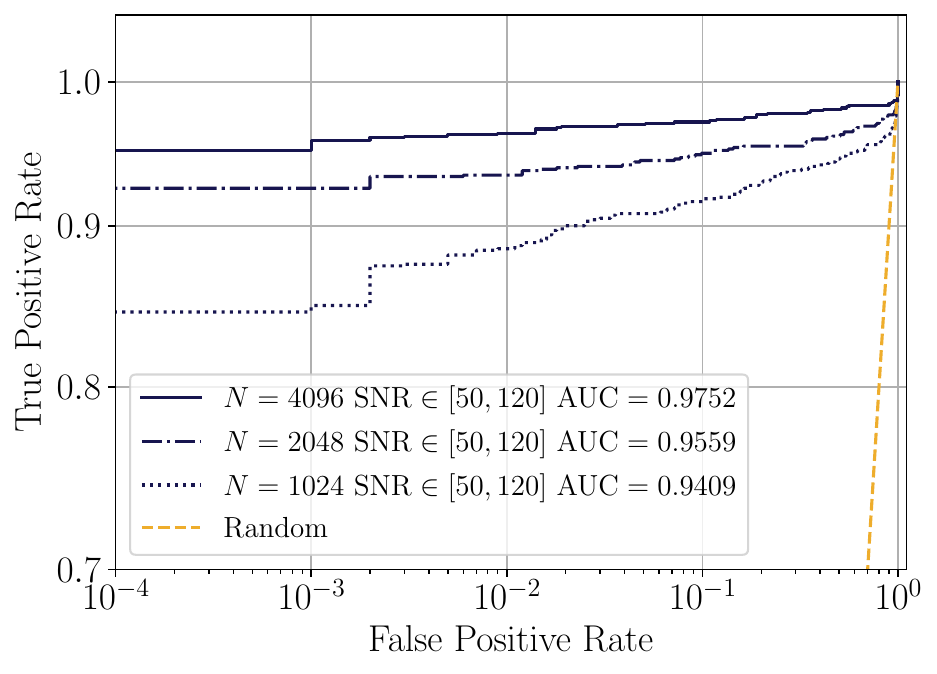}
    }
    \hfill
    \subfloat[\label{fig:roc2}]{%
        \includegraphics[width=0.32\textwidth]{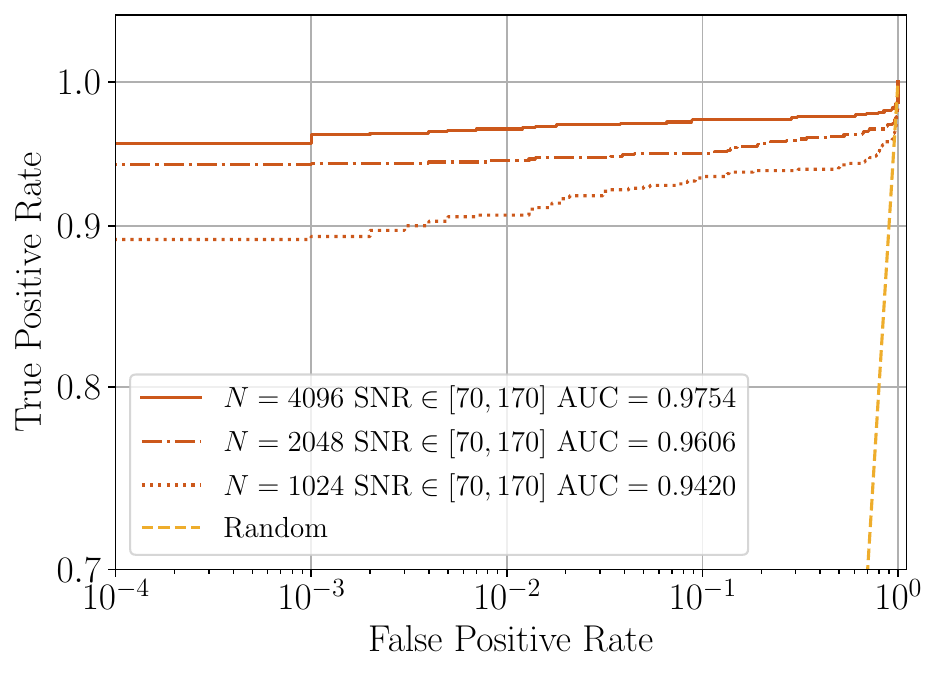}
    }
    \hfill
    \subfloat[\label{fig:roc3}]{%
        \includegraphics[width=0.32\textwidth]{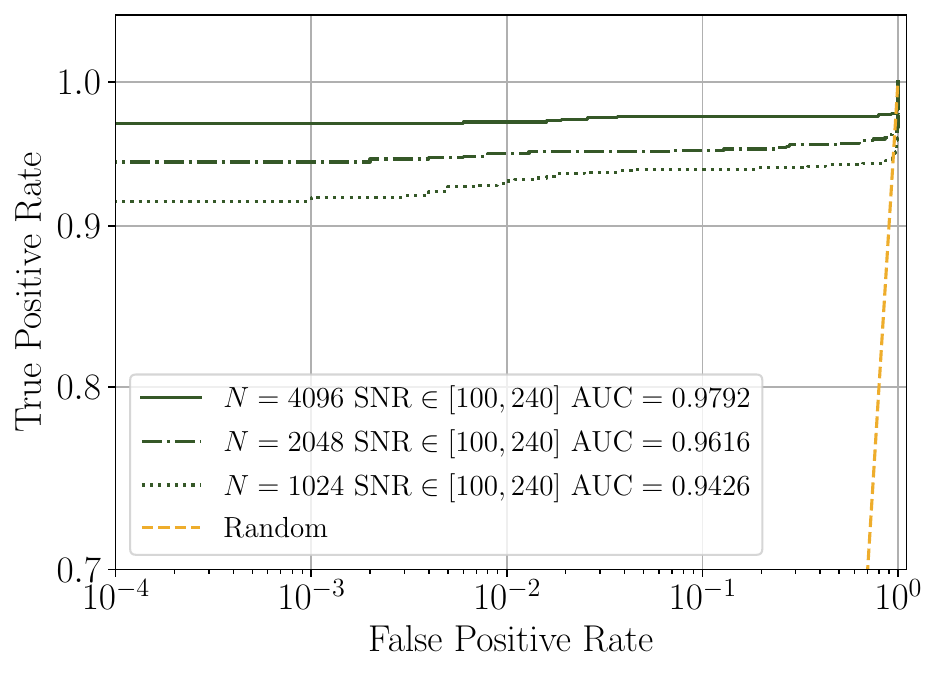}
    }
    \caption{\textbf{\ac{EMRI} detection performance across \ac{SNR} and $\bm{N}$.} All sub-plots depict \ac{ROC} curves for distinct input sample lengths $N$ within specific \ac{SNR} ranges, presented on a logarithmic scale. Each line style signifies the balance between \ac{TPR} and \ac{FPR} for a given sample length, with the area beneath each curve representing the model's efficacy. A reference yellow dashed line indicates the random prediction. The use of logarithmic scales enhances the visibility of performance difference, especially at lower \ac{FPR} levels. (a), Evaluation for $\mathrm{SNR} \in [50, 120]$. (b), Evaluation for $\mathrm{SNR} \in [70, 170]$. (c), Evaluation for $\mathrm{SNR} \in [100, 240]$.}
    \label{fig:3roc}
\end{figure*}

\subsection{DECODE}
In this work, we introduce the \ac{DECODE}, a novel architecture for sequence
modeling tasks, as illustrated in \Cref{fig:network}. The \ac{DECODE} is
inspired by the \acs{TCN} architecture \cite{Bai_2018}, which has been shown to
outperform traditional recurrent architectures across a diverse range of tasks
and datasets. The \ac{DECODE} architecture leverages the strengths of
convolutional networks, which have been proven to be highly effective for
sequence modeling. It incorporates dilated convolutions, which are a powerful
tool for capturing long-range dependencies in sequence data. The causal nature
of the \ac{DECODE} ensures that the model's output at each step is conditioned
on all previous steps, making it suitable for tasks that require an
understanding of sequential dependencies. While the \ac{TCN} and other sequence
modeling architectures have predominantly been applied to time series data, the
\ac{DECODE} stands out in its application to frequency domain data. Detecting
\ac{EMRI} in the time domain presents challenges due to the extended duration
of the signals and their low \ac{SNR}. As illustrated in \Cref{fig:td-data},
the amplitude of the signal is typically three orders of magnitude lower than
the noise, and the data spans a full year. However, as shown in
\Cref{fig:fd-data}, in the frequency domain, the signal's \ac{PSD} has lots of
peaks, with some even reaching the noise level. Despite this shift from time to
frequency domain, the core principles of sequence modeling remain applicable.
The \ac{DECODE} effectively exploits these principles, achieving notable
performance in \ac{EMRI} signal detection.

\paragraph{Causal Sequence Modeling}
The \ac{DECODE} framework is designed for sequence modeling, with a focus on
maintaining causality throughout its structure. Central to \ac{DECODE}'s design
are two fundamental principles. Firstly, the architecture ensures that the
output sequence's length aligns with the input sequence. This alignment is
achieved via a 1D-convolutional network design, where each hidden layer matches
the length of the input layer. To maintain this length consistency, zero
padding of length $(\text{kernel size} - 1)$ is applied. Following this, the
architecture emphasizes the causality of the sequence. This is achieved by
using causal convolutions, which ensure that the output at a particular time
step is convolved only with preceding elements in the previous layer.

\paragraph{Dilated Convolution}
Incorporated into the \ac{DECODE} architecture, dilated convolutions play a
pivotal role in capturing long-range dependencies in sequence data. Drawing
inspiration from the WaveNet \cite{oord_2016}, the \ac{DECODE} employs dilated
convolutions to exponentially expand the receptive field without a significant
increase in computational complexity or number of parameters. We provide an
illustration in \Cref{fig:wavenet}, More formally, for a 1-D sequence input
$\mathbf{x} \in \mathbb{R}^n$ and a filter $f : \{0, . . . , k-1\}
    \longrightarrow \mathbb{R}$, the dilated convolution operation $F$ on element
$s$ of the sequence is defined as:
\begin{align}
    F(s)=\left(\mathbf{x} *_d f\right)(s)=\sum_{i=0}^{k-1} f(i) \cdot \mathbf{x}_{s-d \cdot i} \qc
\end{align}
where $d$ is the dilation factor, $k$ is the filter size (i.e. kernel size), and $s-d \cdot i$ accounts for the direction of the past. When $d = 1$, a dilated convolution reduces to a regular convolution. By employing larger dilations, the receptive field of a \ac{DECODE} is effectively expanded, allowing it to capture long-range dependencies within the sequence data more effectively.

\paragraph{Residual Connections}
Residual connections, another key feature of the \ac{DECODE} architecture, are
designed to facilitate the training of deep networks. These connections,
introduced by He et al. \cite{he_2016}, allow the gradient to flow directly
through the network, mitigating the problem of vanishing gradients that often
jeopardize deep networks. In the \ac{DECODE}, a residual block is composed of
two dilated causal convolutional layers, with a residual connection skipping
over them. If we denote the input to the residual block as $\mathbf{x}$, the
output of the block, $\mathbf{y}$, can be computed as:
\begin{align}
    \mathbf{y} = \operatorname{Activation}(\mathbf{x} + \mathcal{F}(\mathbf{x})) \qc
\end{align}
where $\mathcal{F}(\mathbf{x})$ represents the transformations performed by the dilated causal convolutional layers. This design choice has been shown to improve the performance of deep networks and is a key component of the \ac{DECODE} architecture. The residual block used in the \ac{DECODE} model is illustrated in \Cref{fig:res-block}. Each block comprises two layers of dilated causal convolution, followed by the \ac{ReLU} activation function. Weight normalization \cite{salimans_2016} and dropout \cite{srivastava_2014} are incorporated after each dilated convolution within the residual block.

\paragraph{Loss Function}
In our \ac{DECODE} model, the output of the residual block has a shape of $(H,
    N)$, where $H$ represents the hidden size of our model and $N$ is the length of
the input sequence. The last column of this output is then passed through a
linear layer to generate the predicted probability for \ac{EMRI} signal
detection. To train the model, we use the cross-entropy loss, a common choice
for classification tasks. One of the advantages of using the cross-entropy loss
is its ability to accelerate convergence during training, especially when
compared to other loss functions like mean squared error \cite{bishop_2016}.
The cross-entropy loss for a binary classification problem is given by:
\begin{align}
    \mathcal{L} = -\frac{1}{n} \sum_{i=1}^{n}y_i \log(\mathcal{P}_i) + (1-y_i) \log(1-\mathcal{P}_i) \qs
\end{align}
In this equation, $y_i$ denotes the actual label, while $\mathcal{P}_i$ is the predicted probability for the $i$-th sample, with $n$ representing the total number of samples in the training dataset. The cross-entropy loss quantifies the divergence between the actual label and the predicted probability.

\subsection{Implementation Detail}
% detail setting of model training, software and hardware, memory consuming
For waveform generation of training data, we employed

\texttt{FastEMRIWaveform}\footnote{\url{https://github.com/BlackHolePerturbationToolkit/FastEMRIWaveforms}} \cite{chua_2021,katz_2021} for \ac{EMRI} signal creation and

\texttt{lisa-on-gpu}\footnote{\url{https://github.com/mikekatz04/lisa-on-gpu}} \cite{katz_2022a} for GPU-accelerated detector response simulations, which includes \ac{TDI}. We also integrated additional functionalities from the \texttt{SciPy} library. Our \ac{DECODE} architecture consists of 10 residual blocks, each with a kernel size of 3 and a hidden size of 128. Developed using the \texttt{PyTorch} framework, known for its computational efficiency and speed, computations were performed on a high-performance computing cluster equipped with NVIDIA Tesla V100 GPUs. The training utilized the Adam optimizer with a learning rate of $2\times 10^{-4}$ and a batch size of 64.

\begin{figure*}[htbp]
    \subfloat[\label{fig:tpr-snr}]{%
        \includegraphics[width=0.325\textwidth]{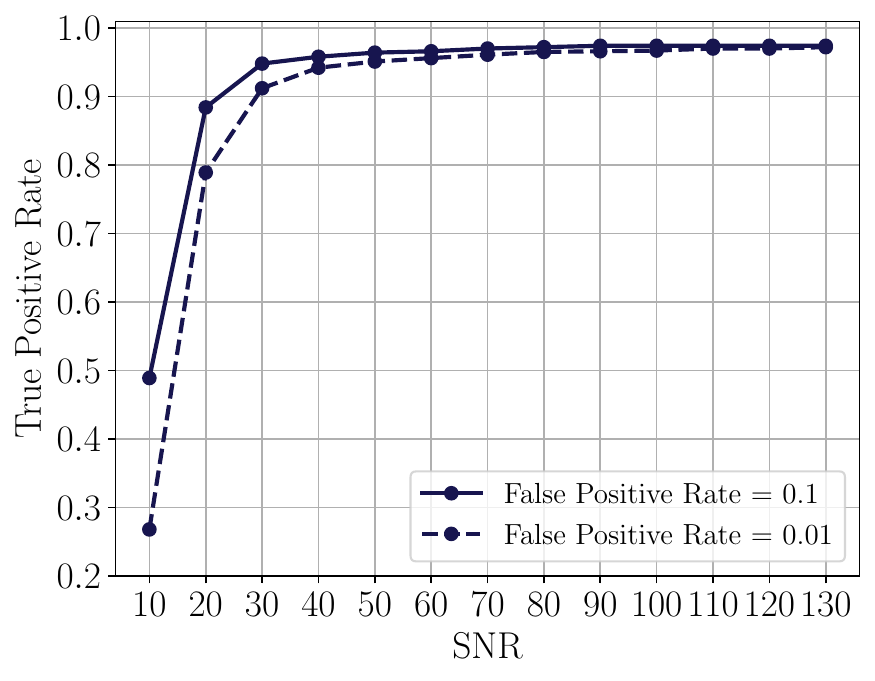}
    }
    \hfill
    \subfloat[\label{fig:tpr-D}]{%
        \includegraphics[width=0.32\textwidth]{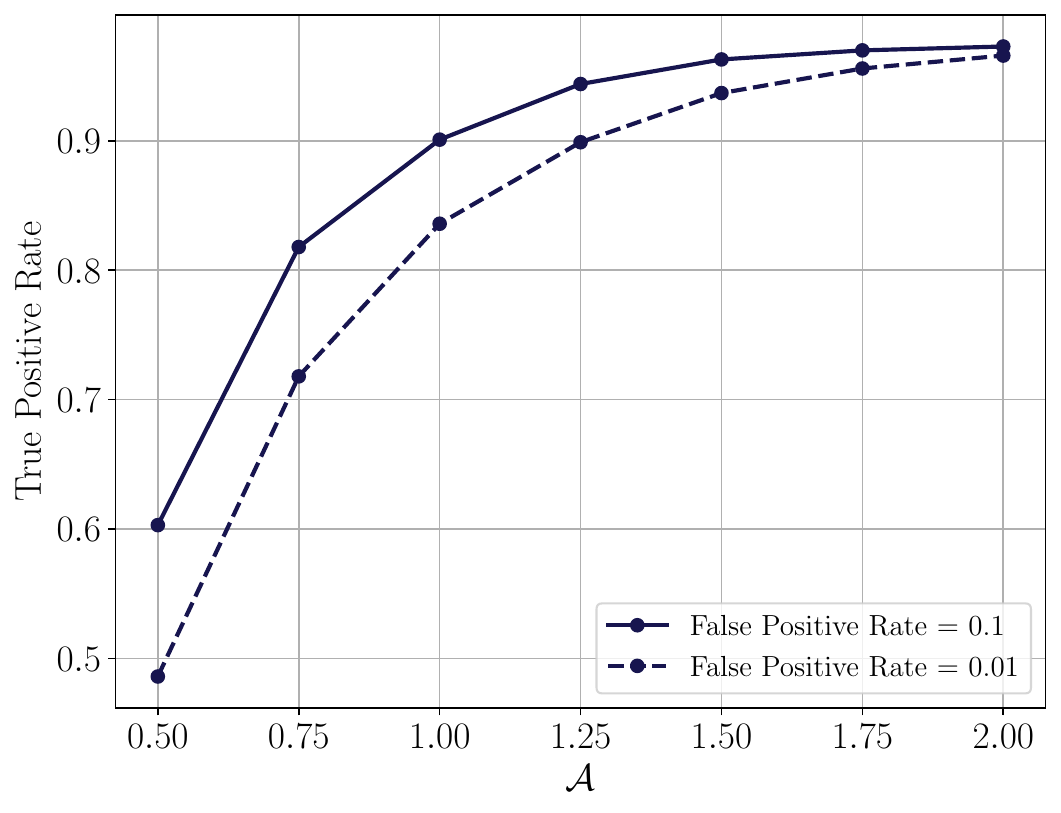}
    }
    \subfloat[\label{fig:tpr-a}]{%
        \includegraphics[width=0.32\textwidth]{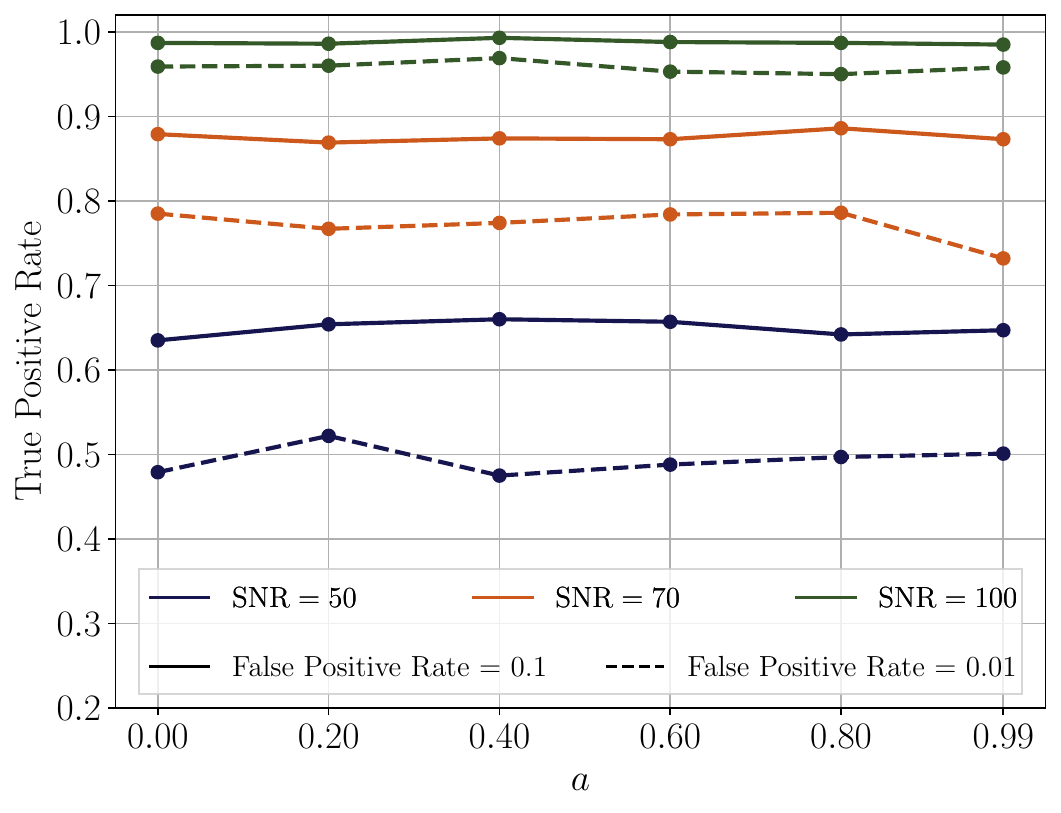}
    }
    \caption{\textbf{Detection capability of \ac{DECODE} across various parameters.} (a), Illustrates the \ac{TPR} as a function of \ac{SNR}, highlighting the model's capability to detect signals with varying strengths. (b) Showcases the \ac{TPR} plotted against the relative amplitude $\mathcal{A}$ (defined in \cref{eq:amp}), emphasizing the model's ability to detect power excesses in the frequency domain and detect signals even when they are submerged within the noise. (c) Explores the \ac{TPR} in relation to the spin parameter $a$, keeping the \ac{MBH} mass consistent at $10^6 M_{\odot}$. This sub-figure is evaluated at three distinct \ac{SNR} levels: 50, 70, and 100, shedding light on the relationship between spin parameters and detection capabilities.}
    \label{fig:intrinsic-1d}
\end{figure*}

\section{\label{sec:result} Results}
\subsection{EMRI Detection Proficiency}
\Ac{ROC} curve and the \ac{AUC} are essential tools for evaluating the performance of models in binary classification tasks. In the context of our study, where the task is to detect \ac{EMRI} signals buried in noise, these tools provide valuable insights. The \ac{ROC} curve, which plots the \ac{TPR} against the \ac{FPR}, offers a visual representation of the model's performance across various threshold settings.  The \ac{AUC}, on the other hand, provides a single, overall measure of the model's performance across all thresholds. A model with perfect discrimination has an \ac{AUC} of 1, while a model performing no better than random guessing has an \ac{AUC} of 0.5.

In our research, we employ \ac{ROC} curves as the primary benchmark to quantify
the performance of the \ac{DECODE}. Our test dataset used here is generated
like the training datasets, i.e. the waveform parameters are uniformly
distributed as shown in \Cref{tab:emri_par} but with different \ac{SNR} range.
As depicted in \Crefrange{fig:roc1}{fig:roc3}, we show three separate \ac{ROC}
curves, with each corresponding to a unique input sample length fed into the
\ac{DECODE}. For the specified input lengths of $N = (1024, 2048, 4096)$, the
\ac{SNR} ranges are set at $[50, 120]$, $[70, 170]$, and $[100, 240]$. The
associated \ac{AUC} values, detailed within the figures, offer quantitative
insight into the model's sensitivity in detecting \ac{EMRI} signals. For
clarity in visual representation, especially at lower \ac{FPR} values,
\Cref{fig:3roc} adopt a logarithmic scale for their axes.

It's noteworthy that our test dataset comprises signals with a duration of 1
year, achieving twice the \ac{SNR} compared to the 3-month data scenario
presented in Ref. \cite{zhang_2022}. While their study tested models on
datasets with $\mathrm{SNR} \in [50, 120]$, we evaluated ours on datasets with
$\mathrm{SNR} \in [100, 240]$. Both datasets, when rescaled for a 1-year
duration, maintain equivalent \ac{SNR} values, implying consistent signal
amplitudes. Impressively, our model attains a \ac{TPR} of 97.5\% at a \ac{FPR}
of 1\% as showcased in \Cref{fig:roc3}.

One significant advantage of deep learning methods over matched filtering-based
approaches is their speed. Once trained, the model can be rapidly deployed for
inference. In our tests, conducted on a single NVIDIA Tesla V100 GPU, our model
processed 2000 data samples in approximately 4 seconds, amounting to less than
$10^{-2}$ seconds per sample.

\subsection{EMRI Detection Efficacy}
In \Cref{fig:intrinsic-1d}, we provide a detailed examination of the
\ac{DECODE}'s performance across different physical parameters.
\Cref{fig:tpr-snr} illustrates the relationship between \ac{TPR} and \ac{SNR}.
The sub-figure clearly demonstrates that as the \ac{SNR} increases, the
\ac{TPR} increases correspondingly, particularly at the specified \ac{FPR}
thresholds of 0.10 and 0.01.

To gain a deeper understanding of the sensitivity of our model, we introduce
the relative amplitude, denoted as $\mathcal{A}$. It is defined as:
\begin{align}
    \label{eq:amp}
    \mathcal{A} = \max_{i \in {A, E}}\sqrt{\frac{S_h^i(f)}{S_n(f)}} \qc
\end{align}
where $S_h^i$ represents the Welch \ac{PSD} of waveform $h_i$. This metric effectively captures the signal's amplitude in the frequency domain. \Cref{fig:tpr-D} plots the \ac{TPR} against the relative amplitude, at \acp{FPR} of 0.1 and 0.01, this sub-figure presents the model's proficiency in discerning power exceeds in the frequency domain. Notably, the \ac{DECODE} can also detect signals that are entirely submerged within the noise.

\captionsetup[subfigure]{labelformat=empty, captionskip=-10pt}
\begin{figure*}[htbp]
    \subfloat[\label{fig:act-map-aak}]{\raisebox{0.2\textwidth}{(a)}\hspace{-0.1cm}
        \includegraphics[width=0.99\textwidth]{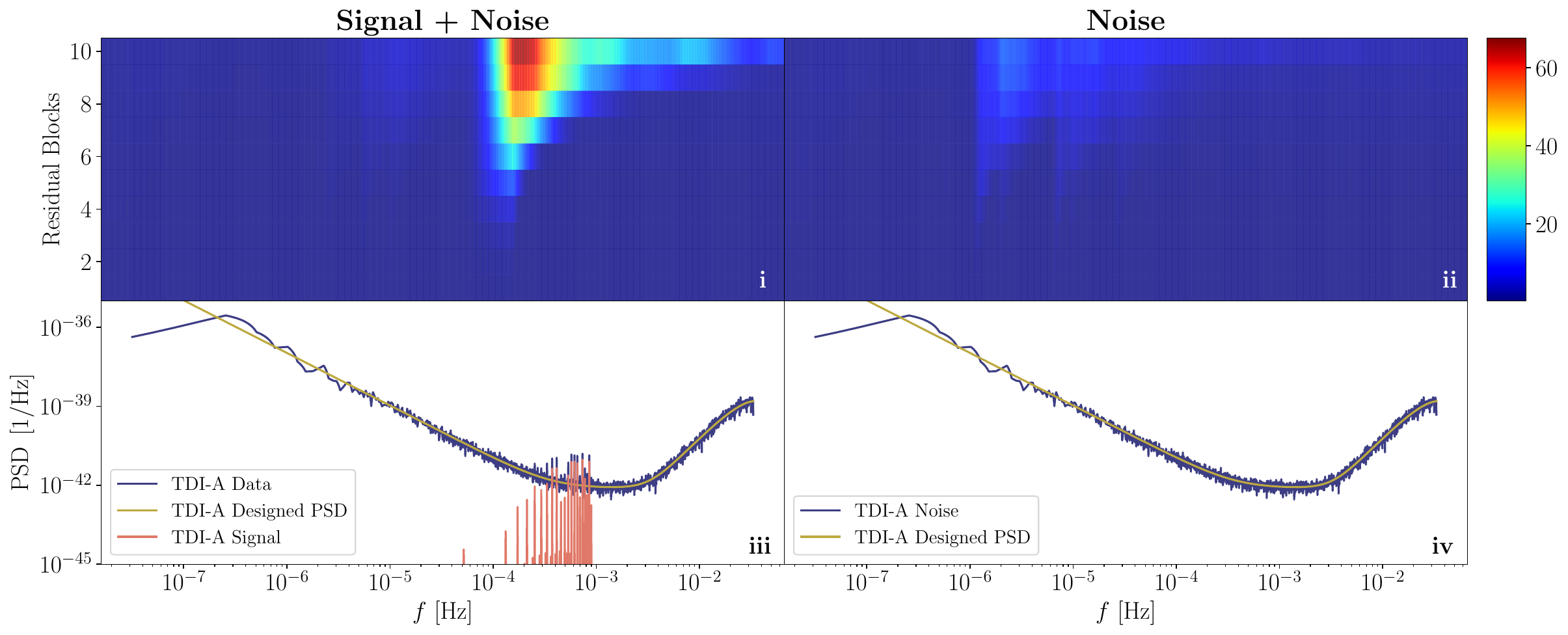}
    }\\
    \vspace{-1.2em}
    \subfloat[\label{fig:act-map-ak}]{\raisebox{0.2\textwidth}{(b)}\hspace{-0.1cm}
        \includegraphics[width=0.99\textwidth]{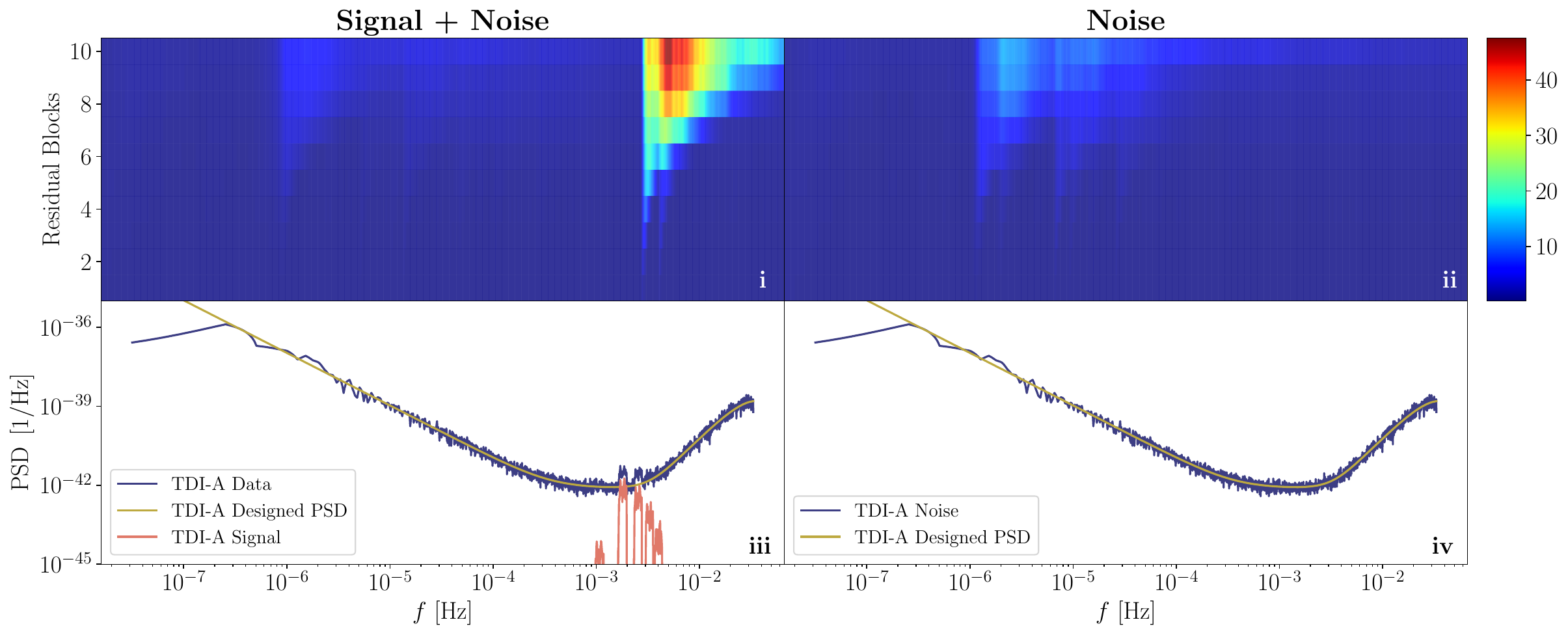}
    }\\
    \vspace{-1.2em}
    \subfloat[\label{fig:act-map-xspeg}]{\raisebox{0.2\textwidth}{(c)}\hspace{-0.1cm}
        \includegraphics[width=0.99\textwidth]{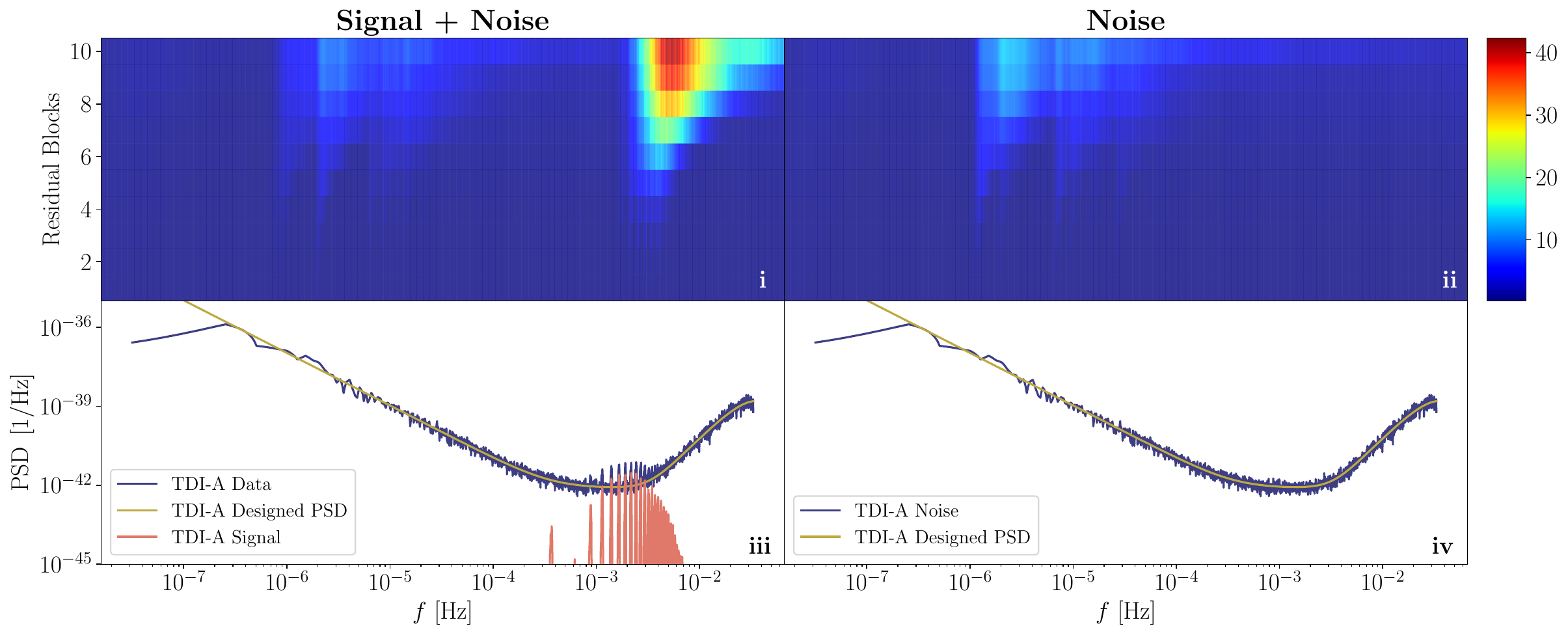}
    }
    \vspace{-1em}
    \caption{\textbf{Interpretability and generalization ability showcase.}
        This figure provides an in-depth visualization of the intermediate outputs from each residual block, demonstrating the model's capability for feature extraction within the frequency domain and it's generalization ability to different waveform templates and gravitational theories.
        For each sub-figure, panels \textbf{i} and \textbf{ii} represent the intermediate results corresponding to the input data samples shown in panels \textbf{iii} and \textbf{iv}. In contrast to the faint activations in panel \textbf{ii}, the noticeable activated neurons in panel \textbf{i} indicate the extraction of essential characteristics when a signal is present in the input. (a), AAK waveform. (b), AK waveform. (c), XSPEG waveform.
    }
    \label{fig:activation-map}
\end{figure*}

In \Cref{fig:tpr-a}, we evaluate the \ac{DECODE}'s sensitivity to varying spin
parameters, while keeping the \ac{MBH} mass constant at $10^6 M_\odot$. The
evaluation, performed at \ac{SNR} levels of 50, 70, and 100 and \ac{FPR}
thresholds of 0.1 and 0.01, indicates that the model's detection performance is
mainly influenced by the \ac{SNR}. In contrast, the spin parameter appears to
have a limited effect on detection, suggesting that the spin parameter
contribution to the overall strength of the \ac{EMRI} signal is relatively
minor.

\subsection{Interpretability}
\ac{CNN}-based models are powerful tools for pattern recognition and prediction. Their unique architecture and operational mechanism make them inherently interpretable, a feature that is particularly valuable in interdisciplinary research. \ac{CNN}-based models learn hierarchical patterns in the data through their convolutional layers, with each layer extracting a set of high-level features from the input data. These features are then used by subsequent layers to understand more complex patterns. This transparent process of feature extraction can be visualized, providing insight into how the network interprets the data and makes predictions.

The activation maps, often used in the context of neural networks, provide a
visual representation of the features that the model identifies and emphasizes
during its processing. Essentially, they capture the output values or
"activations" from various layers or blocks within the network when presented
with an input. These maps offer insights into which parts of the input data the
model finds significant or relevant for a particular task. In the case of the
\ac{DECODE}, the activation maps generated at the output of each residual block
reveal how the model processes and interprets the frequency-domain data of
\ac{EMRI} signals.

The activation maps illuminate the interpretability of the \ac{DECODE}. By
analyzing the outputs of multiple residual blocks, the processes of feature
extraction are made transparent. \Cref{fig:activation-map} provides a detailed
visualization of these maps, demonstrating the ability of the \ac{DECODE} to
distinguish \ac{EMRI} signals from noise. Specifically, panel \textbf{i} of
each sub-figure depicts activation maps for inputs with an \ac{EMRI} signal,
while panel \textbf{iii} depicts the corresponding frequency domain data. These
maps emphasize activated neurons in regions that correspond to the frequency
components of the signal. In contrast, panel \textbf{ii} of each sub-figure
depicts diminished activations for noise-only samples. The corresponding
frequency domain data for these samples is presented in panel \textbf{iv},
validating the model's ability at identifying \ac{EMRI} signals.

\subsection{Generalization Ability}
Generalization ability is the capacity of a model trained on a specific dataset
to perform well on new, untrained data. It indicates how well a model can
extrapolate from its training data to make accurate predictions on unknown
data. In practical applications, a model will frequently be presented with data
that differs from its training set, so this ability is crucial. A model that
generalizes well is robust and flexible, ensuring that it does not simply
memorize the training data but rather understands inherent patterns and
relationships.

In \Cref{fig:act-map-ak} and \ref{fig:act-map-xspeg}, we provide evidence of
the generalization capabilities of our model. Even though the model was only
trained on AAK waveform datasets, it identified the AK waveform accurately
during evaluation with the output probability equal to 1, demonstrating its
ability to generalize across various waveform templates. In contrast, the
model's successful detection of the XSPEG waveform \cite{xin_2019,zhang_2021},
which was formulated using the KRZ metric, demonstrates its generalization
ability with respect to various gravitational theories. These results
demonstrate the generalization ability of the model, suggesting that it is
capable of handling scenarios beyond its training datasets.

\section{\label{sec:discussion} Conclusion and Discussion} The detection of \acp{EMRI} in gravitational wave
astronomy presents a formidable challenge. In this paper, we introduce the
\ac{DECODE}, a state-of-the-art end-to-end \ac{DNN} model designed for the
detection of \ac{EMRI} signals in the frequency domain. By leveraging dilated
causal convolutional layers, the \ac{DECODE} efficiently processes year-long
data. Our evaluations on synthetic datasets have revealed the model's
robustness and efficiency, achieving remarkable detection rates at varied
\ac{SNR} levels. Furthermore, the model's rapid inference capabilities and its
ability to generalize beyond its training parameters but there is still room
for future advancement.

The precision of the EMRI detection model is intrinsically related to the
precision of the training data. While our current training dataset employs the
\ac{TDI}-1.5 detector response, future developments could benefit from the
incorporation of more sophisticated simulations, such as the \ac{TDI}-2.0
technique. This would provide a more accurate simulation of the detector's
response, potentially enhancing the model's applicability in the actual world.

Our current approach primarily focuses on the amplitude information of the
\ac{EMRI} signals. However, the phase information, which has been largely
wasted in this research, holds considerable potential. By integrating
phase-related features into the model, we could capture more intricate patterns
and details of the \ac{EMRI} signals. This may lead to improved detection rates
and lower false alarm rates.

In conclusion, \ac{DECODE} is a step forward in \ac{EMRI} detection. Even
though there are avenues for improvement, its foundational accomplishments
demonstrate its potential as a tool for future space-based \ac{GW} data
analyses.

\begin{acknowledgments}
    The research was supported by the Peng Cheng Laboratory and by Peng Cheng Laboratory Cloud-Brain. This work was also supported in part by the National Key Research and Development Program of China Grant No.~2021YFC2203001 and in part by the NSFC (No.~11920101003 and No.~12021003). Z.C was supported by the ``Interdisciplinary Research Funds of Beijing Normal University'' and CAS Project for Young Scientists in Basic Research YSBR-006.
\end{acknowledgments}

\bibliographystyle{apsrev4-2}
\bibliography{ref}

\end{document}